\documentstyle{article}

\def\sww{\sin^2\theta_W}

\begin{document}

\begin{flushright}
{\Large LAPTH--884/01}
\end{flushright}

\begin{center}
{\Large \bf CompHEP/SUSY package} \\

\vspace{4mm}

Andrei Semenov\\
Laboratoire de Physique Th\'eorique LAPTH\\
 Chemin de Bellevue, B.P. 110, F-74941 Annecy-le-Vieux, Cedex, France.\\[4mm]
Laboratory of Particle Physics,
     Joint Institute for Nuclear Research, \\
141 980 Dubna, Moscow Region, Russian Federation\\[4mm]

\end{center}


\begin{abstract}
The CompHEP software package allows the evaluation of cross section 
and decays of elementary particles
with a high level of automation. Arbitrary tree level
processes can be calculated starting from the set of vertices
prescribed by a given physical model.
This note describes the details of the Minimal Supersymmetric
Standard Model (MSSM)  implementation in the CompHEP
package, and the notation for the particles and parameters of
the MSSM in CompHEP.
\end{abstract}

\section{Introduction}

The CompHEP package \cite{comphep} allows
 to generate automatically the Feynman diagrams for
a process describing collisions or decays of elementary particles starting
from the Feynman rules. CompHEP performs symbolic calculation of the 
squared matrix element and numerical integration over multiparticle phase 
space for the physical process. The goal of the CompHEP/SUSY package
is to develop the description of the Minimal Supersymmetric Standard
Model (MSSM) and some of its extensions in the CompHEP format.

We have used the description of the MSSM physical spectrum and the Lagrangian 
for 3 generations of quarks and leptons given in \cite{rosiek}.
The LanHEP software package \cite{lhep} was used to generate the
Feynman rules for the MSSM Lagrangian. LanHEP  allows to write 
the Lagrangian in terms of initial fields (multiplets) before 
$SU(2)\times U(1)$ breaking, to use the superpotential formalism and the 2-component 
spinor notation. Thus the LanHEP input model description is compact and
close to textbook formulae.
LanHEP outputs the Feynman rules as CompHEP
tables describing the physical model 
(a LaTeX output is also possible). The Feynman rules were successfully
compared with the ones listed in \cite{rosiek}, as well as with many results
of other publications. The comparision to GRACE/SUSY program
was successfully done for many processes.

We have used the notation of independent 
parameters same as in the ISASUSY package \cite{ISASUGRA} 
which is more popular than one of \cite{rosiek} (we give explicit mass
matrices 
below). We also add the effective Higgs potential to the tree
level MSSM Lagrangian to incorporate the radiative corrections to
the masses and couplings of Higgs bosons in a gauge invariant way.

Several CompHEP/SUSY models can be found on the WWW page

{\tt http://theory.sinp.msu.ru/\~{}semenov/mssm.html} 

The basic CompHEP/SUSY implementation of MSSM assumes massless
fermions of the 1st and 2nd generation, and 
left/right-handed states for the corresponding sfermions.
Stops, sbottoms and staus are mass eigenstates, the mixing
due to fermion masses and trilinear soft coupling constants
is taken into account in the calculation of the sfermion masses
and interaction vertices. In the existing extension of the basic CompHEP/SUSY 
model,
the fermion masses and sfermions mixing of the 2nd 
generations are also taken into account.
We neglect the mixing of the different flavour left-handed squarks
due to off-diagonal soft masses and Cabbibo-Kobayashi-Maskawa (CKM) 
matrix elements\cite{rosiek}.
To keep our model explicitely gauge invariant, we assume
that only the Cabbibo angle contribute to the  CKM matrix, and
that the soft masses for the 1st and 2nd sfermion generations are equal.

Besides the basic MSSM, there is a CompHEP/SUSY model with R-parity violation. 
Some other extensions will be described in forthcoming publications.

\section{Model parameters}

First, the user has access to the Standard Model parameters.
We list those parameters with their default values.
Some parameters, like s-, c-quark masses, CKM parameters do not appear in 
the basic CompHEP MSSM
since we neglect them. (Note that the default values might be
changed in future versions).

\vspace{.5cm}
\begin{tabular}{lll}
\hline\hline 
{\tt EE }   &0.31223      &Electromagnetic coupling constant at $M_Z$ \\ 
{\tt GG  }   &1.238        &Strong coupling constant at $M_Z$ \\ 
{\tt SW  }   &0.473        &sinus of the Weinberg angle, $\sin\theta_W$ \\ 
{\tt s12 }   &0.221        &Parameter of CKM matrix (sinus of Cabbibo angle) \\ 
{\tt s23 }   &0.041        &Parameter of CKM matrix \\ 
{\tt s13 }   &0.0035       &Parameter of CKM matrix \\ 
{\tt MZ }    &91.1884      &mass of $Z$ boson\\ 
{\tt Mm  }   &0.1057       &mass of muon\\ 
{\tt Mtau }    &1.777      &mass of tau-lepton\\ 
{\tt Mc }    &1.42         &mass of c-quark\\ 
{\tt Ms }    &0.2          &mass of s-quark\\ 
{\tt Mtop }  &175          &mass of t-quark\\ 
{\tt Mb  }   &4.62         &mass of b-quark\\ 
{\tt wtop }  &1.7524       &width of t-quark\\ 
{\tt wZ }    &2.4944       &$Z$ boson width\\ 
{\tt wW }    &2.08895      &$W$ boson width\\ 
\hline\hline 
\end{tabular} 
\vskip .5cm

The following parameters are specific for the MSSM, they describe essentially
MSSM extended Higgs sector and soft supersymmetry breaking potential.
Our conventions correspond to the ISASUSY notation.
\vskip .5cm
\begin{tabular}{lll}
\hline \hline 
{\tt tb }  &   $\tan\beta$ & Ratio of the vacuum expectation values of the Higgs doublets \\   
{\tt MH3 } & $M_A$ & Mass of the pseudoscalar Higgs\\   
{\tt mu  } &$\mu$  & Higgs mass parameter \\ 
{\tt MG1 } &$M_1$  & $U(1)$ gaugino mass\\ 
{\tt MG2 } & $M_2$ & $SU(2)$ gaugino mass\\ 
{\tt MG3 } & $M_3$ & Gluino mass\\    
{\tt Ml2 } & $M_{\tilde{L}_{2L}}$ & Left-handed slepton mass, 2nd generation \\ 
{\tt Ml3 } & $M_{\tilde{L}_{3L}}$ & Left-handed slepton mass, 3rd generation \\ 
{\tt Mr2  }& $M_{\tilde{\mu}_R}$ & Right-handed slepton mass, 2nd generation\\ 
{\tt Mr3 } & $M_{\tilde{\tau}_R}$ & Right-handed slepton mass, 3rd generation\\  
{\tt Mq2 } & $M_{\tilde{Q}_{2L}}$ & Left-handed squark mass, 2nd generation\\ 
{\tt Mq3 } & $M_{\tilde{Q}_{3L}}$ & Left-handed squark mass, 3rd generation\\     
{\tt Mu2 } & $M_{\tilde{c}_R}$ & Right-handed u squark mass, 2nd generation\\ 
{\tt Mu3 } & $M_{\tilde{t}_R}$ & Right-handed u squark mass, 3rd generation\\  
{\tt Md2 } & $M_{\tilde{s}_R}$ & Right-handed d squark mass, 2nd generation\\ 
{\tt Md3 } & $M_{\tilde{b}_R}$ & Right-handed d squark mass, 3rd generation\\ 
{\tt Atop } & $A_t$ & Trilinear coupling for top quark \\   
{\tt Ab } & $A_b$ & Trilinear coupling for b-quark \\   
{\tt Atau }&  $A_\tau$ & Trilinear coupling for $\tau$-lepton\\   
\hline \hline
\end{tabular} 
\vskip .5cm

We assume here that soft masses for the first generation coincide with 
 the parameters of the second generation, but in future extensions
 the parameters {\tt Ml1, Mr1, ...} may appear.

The derived parameters needed for a given process such as physical masses, mixings and so on
are computed from those listed above and their values
can be found in the CompHEP {\it Constraints} menu.

\section{Model particles}

The notation for gauge bosons are the same as in the CompHEP Standard Model: 
{\tt A}, {\tt Z}, {\tt W+}, {\tt W-}, {\tt G}. The gluino is named as
{\tt \~{}G} (with mass {\tt MG3}). The notation for other particles follows
with more details.
 
\subsection{Charginos}

Two charginos $\chi_1^+$, $\chi_2^+$ are mass eigenstates resulting from
mixing of charged wino and higgsino. CompHEP names for these two particles
are {\tt \~{}1+}, {\tt \~{}2+}. Two mixing matrices are involved to diagonalize
the mass matrix of charginos:

$$
\left(\begin{array}{cc} 
{\rm \tt Zm11} & {\rm \tt Zm21} \\ 
{\rm \tt Zm12} & {\rm \tt Zm22} \\ 
\end{array}\right)
\left(\begin{array}{cc} 
M_2 & \sqrt{2}M_W\sin\beta\\ 
\sqrt{2}M_W\cos\beta &   \mu     \\ 
\end{array}\right)
\left(\begin{array}{cc} 
{\rm \tt Zp11} & {\rm \tt Zp12} \\ 
{\rm \tt Zp21} & {\rm \tt Zp22} \\ 
\end{array}\right) = diag({\rm \tt MC1},{\rm \tt MC2}).
$$

The values of the chargino masses {\tt MC1}, {\tt MC2}, and mixing elements {\tt Zm\it ij},
{\tt Zp\it ij} can be found in the CompHEP {\it Constraints} menu.

\subsection{Neutralinos}

The four neutralinos $\chi^0_i$ are named  {\tt \~{}o1}, {\tt \~{}o2}, {\tt \~{}o3}, 
{\tt \~{}o4}. The 4x4 mixing matrix $Zn_{ij}$ is defined in the bino-wino-higgsino
basis, so
$$ Z_n^{-1} {\cal M_N} Z_n = diag({\rm \tt MNE1}, {\rm \tt MNE2}, {\rm \tt MNE3}, {\rm \tt MNE4}), $$
where
$$
{\cal M_N}=\left(\begin{array}{cccc}
M_1 & 0 & -M_Z\cos\beta\sin\theta_W & M_Z\sin\beta\sin\theta_W\\
0 & M_2 & M_Z\cos\beta\cos\theta_W & -M_Z\sin\beta\cos\theta_W\\
-M_Z\cos\beta\sin\theta_W& M_Z\cos\beta\cos\theta_W&  0 & -\mu \\
M_Z\sin\beta\sin\theta_W & -M_Z\sin\beta\cos\theta_W& -\mu & 0 \\
\end{array}
\right)
$$

The values of {\tt Zn11}, ... {\tt Zn44} and masses {\tt MNE1 ... MNE4}
are in the CompHEP {\it Constraints} menu. Note that our definition of the mixing
matrix is the inverse of that commonly used, the same is true
for charginos and sfermions.

\subsection{Sleptons}

For each of the 3 lepton generations, there are two charged sleptons and 
one sneutrino. For  the 3rd generation, the stau mixing matrix
is:
$$
{\cal M}_{\tilde{\tau}}^2=\left(\begin{array}{cc}
M_{\tilde{L}_{3L}}^2+m_\tau^2+M_Z^2\cos 2\beta(-{1\over 2} +\sww) & m_\tau(A_\tau-\mu\tan\beta)\\
m_\tau(A_\tau-\mu\tan\beta) &   M_{\tilde {\tau}_R}^2+m_\tau^2-M_Z^2\cos 2\beta\sww
   \\
\end{array}
\right)
$$

The following table summarizes the notation for particles and parameters used in 
the basic CompHEP/SUSY model. 

\tt
\vspace{.5cm}

\begin{tabular}{|l|l|l|l|l|l|l|}\hline
 {\sl flavour } & $e$  & $\mu$ & $\tau$ & $\nu_e$ & $\nu_\mu$ & $\nu_\tau$ \\ \hline 
{\rm fermion} &  e   &  m    &  l     &  ne     & nm        & nl    \\ 
{\rm left scalar} & \~{}eL & \~{}mL &  &  \~{}ne & \~{}nm  &  \~{}nl \\ 
{\rm right scalar} & \~{}eR & \~{}mR &    &   &    &   \\ 
{\rm light scalar }&  &  & \~{}l1 & & & \\ 
{\rm heavy scalar }&  &  & \~{}l2 & & & \\ \hline 
{\rm left scalar mass} & MSeL & MSmL &  &  MSne & MSnmu  &  MSntau \\ 
{\rm right scalar mass} & MSeR & MSmR &  &   &   &   \\ 
{\rm light scalar mass}&  &  & MStau1 & & & \\ 
{\rm heavy scalar mass}&  &  & MStau2 & & & \\ 
{\rm $M_{LL}$ }&  &  & MSlLL & & & \\ 
{\rm $M_{LR}$ }&  &  & MSlLR & & & \\ 
{\rm $M_{RR}$ }&  &  & MSlRR & & & \\ 
{\rm $Z_{1L}$ } &  &  & Zl33 & & & \\ 
{\rm $Z_{2L}$ } &  &  & Zl36 & & & \\ 
{\rm $Z_{1R}$ } &  &  & Zl63 & & & \\ 
{\rm $Z_{2R}$ } &  &  & Zl66 & & & \\ 
{\rm  $\theta$ }&  &  & MSlth & & & \\ 
\hline
\end{tabular}
\vspace{.5cm}
\rm

The parameters $M_{LL}$, $M_{LR}$, and $M_{RR}$ are the elements of the slepton 
squared mass matrix 
to be diagonalized (in $GeV^2$);
the $Z_{iL/R}$ values set the mixing of $s_{L,R}$ left/right states into the $s_{1,2}$ mass
states:
\begin{eqnarray}
s_L& = & s_1Z_{1L}+s_2Z_{2L},\nonumber \\
s_R& = & s_1Z_{1R}+s_2Z_{2R},\nonumber
\end{eqnarray}
and $\theta$ is the mixing angle which allows to represent the latter formulae as
\begin{eqnarray}
s_L& = & s_1\cos\theta-s_2\sin\theta,\nonumber \\
s_R& = & s_1\sin\theta+s_2\cos\theta.\nonumber
\end{eqnarray}
The $\theta$ value is given in  radians.

\subsection{Squarks}

There are two squarks corresponding to each of the 6 quarks of the Standard
Model.  The mass matrices for up and down squarks of
the 3rd generation are:

$$
{{\cal M}_{\tilde{t}}}=\left(\begin{array}{cc}
M_{\tilde{Q}_L}^2+m_t^2+M_Z^2\cos 2\beta({1 \over 2}-\frac{2}{3}\sww) & m_t(A_t-\mu\cot\beta)\\
m_t(A_t-\mu\cot\beta) &   M_{\tilde t_{R}}^2+m_t^2+\frac{2}{3}M_Z^2\cos 2\beta\sww
   \\
\end{array}
\right)
$$

$$
{\cal M}_{\tilde{b}}=\left(\begin{array}{cc}
M_{\tilde{Q}_L}^2+m_b^2+ M_Z^2\cos 2\beta(-{1 \over 2} +\frac{1}{3}\sww)& m_b(A_b-\mu\tan\beta)\\
m_b(A_b-\mu\tan\beta) &   M_{\tilde {b}_R}^2+m_b^2-\frac{1}{3}M_Z^2\cos 2\beta\sww
   \\
\end{array}
\right).
$$

The following table summarizes the  particles and values used in 
the basic model. The notation is similar to the one used for sleptons.

\tt
\vspace{.5cm}

\begin{tabular}{|l|l|l|l|l|l|l|}\hline 
{\sl flavour }    & $d$  & $u$ & $s$  & $c$ & $b$ & $t$ \\ \hline 
{\rm fermion} &  d   &  u    &  s       & c     & b        & t    \\ 
{\rm left scalar} & \~{}dL & \~{}uL & \~{}sL &   \~{}cL &   &   \\ 
{\rm right scalar} & \~{}dR & \~{}uR &  \~{}sR &    \~{}cR &    &   \\ 
{\rm light scalar }&  &   &  & & \~{}b1 & \~{}t1 \\ 
{\rm heavy scalar }&  &   &  & & \~{}b1 & \~{}t2 \\ \hline 
{\rm left scalar mass} & MSdL & MSuL & MSsL &  MScL &  &     \\ 
{\rm right scalar mass}& MSdR & MSuR & MSsR &  MScR &  &  \\ 
{\rm light scalar mass}&      &      &       &  & MSbot1 & MStop1 \\ 
{\rm heavy scalar mass}&      &      &       &  & MSbot2 & MStop2 \\ 
{\rm $M_{LL}$ }&    &   &   &  & MSbLL & MStLL \\ 
{\rm $M_{LR}$ }&    &   &   &  & MSbLR & MStLR \\ 
{\rm $M_{RR}$ }&    &   &   &  & MSbRR & MStRR\\ 
{\rm $Z_{1L}$ } &  &  &  &  & Zd33 & Zu33 \\ 
{\rm $Z_{2L}$ } &  &  &  &  & Zd36 & Zu36 \\ 
{\rm $Z_{1R}$ } &  &  &  &  & Zd63 & Zu63 \\ 
{\rm $Z_{2R}$ } &  &  &  &  & Zd66 & Zu66 \\ 
{\rm $\theta$ } &  &  &  &  & MSbth & MStth\\ 
\hline
\end{tabular}

\rm
 
\vskip .5cm

\subsection{Higgs particles}

The Higgs particles: $h^0$, $H^0$, $H^\pm$, $A^0$ are named in CompHEP as {\tt h, H, H+/H-, H3} 
correspondingly. Their masses are denoted as {\tt Mh, MHH, MHc, MH3}.

It is well known that the masses of Higgs bosons in the
MSSM are significantly shifted by the radiative corrections.
There are several techniques to compute the corrections
to Higgs masses. We use the effective potential 
 (see \cite{haber,higgspot} and
references therein) to keep the explicit gauge 
invariance of the CompHEP model.
The Higgs potential now is an extension
of the MSSM potential:

\begin{eqnarray}
V_{eff} & = & m_{11}^2 (H_1 H_1^+) + m_{22}^2 (H_2 H_2^+)
              - [m_{12}^2 (\epsilon H_1 H_2) + h.c. ] \nonumber \\
	&   & + \frac{1}{2}\big[\frac{1}{4}(g^2+g'^2) + \lambda_1\big]
	                    (H_1 H_1^+)^2
	      +\frac{1}{2}\big[\frac{1}{4}(g^2+g'^2) + \lambda_2\big]
	                    (H_2 H_2^+)^2 \nonumber \\
	&  &  +\big[\frac{1}{4}(g^2-g'^2) + \lambda_3\big]
	                    (H_1 H_1^+)(H_2 H_2^+)
	      +\big[-\frac{1}{2}g^2 + \lambda_4\big]
	                    |\epsilon H_1 H_2|^2\nonumber\\ 
	& &   +\big(\lambda_5 (\epsilon H_1 H_2)^2
	           +\big[ \lambda_6 (H_1 H_1^*) + \lambda_7 (H_2 H_2^*)\big]
		         (\epsilon H_1 H_2) + h.c. \big) \nonumber
\end{eqnarray}

The case of all $\lambda_i$ being zero corresponds to the original tree-level MSSM
potential. 

Here the parameters $m_{11}^2$, $m_{22}^2$ can be fixed by the
condition of potential minimization, and the parameter $m_{12}^2$ 
can be fixed if we choose $M_A$
as independent variable. Then the masses of Higgs bosons and mixing 
angle $\alpha$ can be derived (see \cite{higgspot} for formulae).

Analytical formulae for $\lambda_i$ are given, for example, in \cite{carena}.
The CompHEP notation for $\lambda_i$ is {\tt dlh1 ... dlh7}.

Many authors use different approaches to calculate Higgs masses,
involving diagrammatic calculation rather than effective potential
and RGE. So, many articles and computer programs calculate
only the CP-even Higgs masses rather than the full $\lambda_i$ set.
This relates in particular to the {\tt FeynHiggs} package. 
Note that such a prescription bounds only several of the 7 parameters,
and we have to deal with $\lambda$'s to keep our model gauge invariant.

In the case when the Higgs masses are given,
we derive the values for $\lambda_{1,2,3,4}$ assuming that
$M_H$, $M_h$, $M_{H\pm}$ and $\alpha$ are independent variables.
The masses {\tt Mh, MHH, MHc} and $\sin\alpha$ {\tt sa} become 
free parameters. Then we use the {\tt FeynHiggsFast}
package \cite{fhf} to evaluate these parameters. This
approach allows us to take into account the corrections to the
Higgs masses and also to self-interaction vertices 
(see the discussion in \cite{higgspot}).

\section{Width parameters}

There is a number of parameters representing particle widths.
In the current version the values for these parameters are set by the user. These
parameters can be important in case of s-channel resonant production
of a particle.

\begin{itemize}
\item {\tt wC1, wC2} --- the widths of two charginos;
\item {\tt wNE2, wNE3, wNE4} --- the widths of three neutralinos 
	(except the lightest one);
\item {\tt wSG } --- the  width of gluino;
\item {\tt wSe1, wSe2, wSmu1, wSmu2, wStau1, wStau2} ---
	 the widths of charged sleptons;
\item {\tt wSne, wSnmu, wSntau} --- the  widths of sneutrinos;
\item {\tt wSu1, wSu2, wSd1, wSd2, wSs1, wSs2, wSc1, wSc2, wSbot1, wSbot2, 
	wStop1, wStop2} --- the widths of squarks;
\item {\tt wh } --- width of light CP-even Higgs;
\item {\tt wHh } --- width of heavy CP-even Higgs;
\item {\tt wHc } --- width of charged Higgs;
\item {\tt wH3 } --- width of CP-odd Higgs.
\end{itemize}

%
%
%
%
%
%

\section{SUGRA and GMSB models}

The number of  independent parameters in the MSSM can be reduced 
in the context of the  SUGRA or GMSB models. More specifically,
the parameters of soft SUSY breaking terms (gaugino masses $M_1$, $M_2$, $M_3$,
sparticle masses $M_L$, $M_R$, $M_Q$ ..., trilinear
couplings $A_t$, $A_b$, $A_\tau$ are computed from just a few parameters.

It is possible to link CompHEP/SUSY with the ISASUSY package \cite{ISASUGRA},
which includes SUGRA and GMSB models. For this, Isajet library should be
installed on the computer.
ISASUGRA is used to calculate soft SUSY breaking masses and couplings which 
were CompHEP independent variables. The masses of the particles are 
calculated by CompHEP from these soft parameters as before. In particular,
masses of CP-even Higgs particles are calculated by {\tt FeynHiggsFast} package,
while ISASUSY gives the mass of the CP-odd Higgs boson.

SUGRA option:

\vskip 3mm
\begin{tabular}{|p{5cm}|p{8cm}|} \hline
CompHEP notation & parameters \\ \hline 
{\tt tb} & ratio of vacuum expectation value\\[2mm]
{\tt suM0} & $M_0$ -- universal mass for scalars \\[2mm]
{\tt suMHF} & $M_{1/2}$  universal mass for fermions \\[3mm]
{\tt suA0 } & $A_0$ universal trilinear coupling \\[2mm]
{\tt suMU } & {\tt sign$(\mu)$} - sign of higgsino mass parameter  \\[2mm]
{\tt suMODE } & choose the mSUGRA model  \\
\hline
\end{tabular}
\vskip 4mm

Here {\tt suMODE=1} means default SUGRA calculation,
 {\tt suMODE=2} is SUGRA with true gauge coupling unification at the GUT scale.

GMSB option:

\vskip 3mm
\begin{tabular}{|p{5cm}|p{8cm}|} \hline
CompHEP notation & parameters \\ \hline 
{\tt tb} & ratio of vacuum expectation value\\[2mm]
{\tt gmLAM} &  $\Lambda$ --- SUSY breaking parameter \\[2mm]
{\tt gmMES} &  $M_{MES}$ messenger mass \\[3mm]
{\tt gmN5 } & $N_5$ effective number of messenger generations \\[2mm]
{\tt gmMU } & {\tt sign$(\mu)$} - sign of higgsino mass parameter  \\[2mm]
{\tt gmMODE } & should be 1 (reserved for extensions) \\
\hline
\end{tabular}
\vskip 4mm

\section{Massive fermions of the second generation}

There is an extension of the basic CompHEP/SUSY model which
takes into accont the masses of the fermions of the second
generation and the mixing of the corresponding scalar superpartners.
We introduce the parameters for trilinear couplings {\tt Am, Ac, As}
for muon, c- and s-quarks. The sfermion names and parameters read as:
\tt
\vspace{.5cm}

\begin{tabular}{|l|l|l|l|}\hline 
{\sl flavour }    & $\mu$  & $s$ & $c$   \\ \hline 
{\rm fermion} &  m   &  s    &  c          \\ 
{\rm light scalar }& \~{}m1 &  \~{}s1 & \~{}c1  \\ 
{\rm heavy scalar }& \~{}m2 &  \~{}s2 & \~{}c2  \\ \hline 
{\rm light scalar mass}& MSmu1  & MSs1  &  MSc1  \\ 
{\rm heavy scalar mass}& MSmu2  & MSs2  &  MSc2   \\ 
{\rm $M_{LL}$ }& MSmuLL  & MSsLL & MScLL \\ 
{\rm $M_{LR}$ }& MSmuLR  & MSsLR & MScLR \\ 
{\rm $M_{RR}$ }& MSmuRR  & MSsRR & MScRR\\ 
{\rm $Z_{1L}$ } & Zl22 & Zd22 & Zu22 \\ 
{\rm $Z_{2L}$ } & Zl25 & Zd25 & Zu25 \\ 
{\rm $Z_{1R}$ } & Zl52 & Zd52 & Zu52 \\ 
{\rm $Z_{2R}$ } & Zl55 & Zd55 & Zu55 \\ 
{\rm $\theta$ } & MSmuth & MSsth & MScth\\ 
\hline
\end{tabular}

\rm
 
\vskip .5cm

\section{R-parity violation}

We have implemented in CompHEP the MSSM extension with R-parity
violation. We incorporate here only the trilinear terms breaking the leptonic 
number. The  superpotential reads:
$$
W_{RPV} =  \lambda_{ijk} \epsilon L_i L_j E_k  + \lambda_{ijk}' \epsilon L_i Q_j D_k,
$$
where the indices $i$, $j$, $k$ enumerate 3 generation of matter fields.

The CompHEP notation reads {\tt Ll\it ijk} for $\lambda_{ijk}$ and
{\tt Lq\it ijk} for $\lambda_{ijk}'$.
A lot of processes were compared
with the SUSYGEN package \cite{susygen} with good agreement. The values for 
$\lambda_{ijk}$ should be set twice greater in SUSYGEN than 
in CompHEP due to the different convention.

\section*{Acknowledgements}

This work was supported in part by the CERN--INTAS grant 
99--0377 and by RFFR grant 01-02-16710. G.B\'elanger and F.Boudjema
have compared many processes against GRACE/SUSY and other programs,
some bugs were found and corrected.

\end{document}